\documentclass[11pt,cite,epsfig,psfrag]{article}
\usepackage[utf8]{inputenc}
\usepackage{float}
\usepackage{placeins}
\usepackage{geometry}
 \usepackage{graphicx}
 \usepackage{subfig}
  \usepackage{amsmath} 
  \graphicspath{{Images@hp/}}
  \usepackage[usenames, dvipsnames]{color}
\usepackage{wrapfig}
\usepackage{boxedminipage}
\usepackage{multirow}

\usepackage{fullpage}
\usepackage{amsmath}
\usepackage{amssymb}
\usepackage{graphicx}
\usepackage{mathrsfs}
\usepackage{wrapfig}
\usepackage{boxedminipage}
\usepackage{epsfig}
\usepackage{setspace}
\usepackage{float}
\usepackage{hyperref}
\usepackage{enumerate}
\usepackage{cite}
\usepackage[font=footnotesize,labelfont=rm]{caption}

\usepackage[numbers,sort&compress]{natbib}

\def\be{\begin{equation}}
\def\ee{\end{equation}}
\def\bea{\begin{eqnarray}}
\def\eea{\end{eqnarray}}

\numberwithin{equation}{section}

 \newcommand{\RN}[1]{%
   \textup{\uppercase\expandafter{\romannumeral#1}}%
 }
\begin{document}
\thispagestyle{empty}

\vskip 2cm

\begin{center}
{\Large \bf Topology of critical points and Hawking-Page transition}
\end{center}

\vskip .2cm

\vskip 1.2cm

\centerline{ \bf   Pavan Kumar Yerra \footnote{pk11@iitbbs.ac.in}, Chandrasekhar Bhamidipati\footnote{chandrasekhar@iitbbs.ac.in} and Sudipta Mukherji\footnote{mukherji@iopb.res.in}
}

\vskip 7mm 
\begin{center}{ $^{1},^{3}$ Institute of Physics, Sachivalaya Marg, \\ Bhubaneswar, Odisha, 751005, India \\ and\\Homi Bhabha National Institute, \\Training School Complex, \\Anushakti Nagar, Mumbai, 400085, India}	
\end{center}

\begin{center}{ $^{2}$ School of Basic Sciences\\ 
Indian Institute of Technology Bhubaneswar \\ Bhubaneswar, Odisha, 752050, India}
\end{center}

\vskip 1.2cm
\vskip 1.2cm
\centerline{\bf Abstract}

Using the Bragg-Williams construction of an off-shell free energy we compute the topological charge of the Hawking-Page transition point for black holes in AdS. A computation following from a related off-shell effective potential in the boundary gauge dual matches the value of topological charge obtained in the bulk. We also compute the topological charges of the equilibrium phases of these systems, which follow from the saddle points of the appropriate free energy. The locally stable and unstable phases turn out to have topological charges opposite to each other, with the total being zero, in agreement with the result obtained from a related construction~\cite{Wei:2022dzw} .

\vskip 0.5cm
\noindent

\newpage
\setcounter{footnote}{0}
\noindent

\baselineskip 15pt

Phase transitions and critical phenomena are interesting topics in thermodynamics, and particularly, in the context of black holes.  Black hole thermodynamics~\cite{Bekenstein:1973ur,Bardeen:1973gs,Hawking:1975vcx} allows us a thorough investigation of microscopic degrees of freedom of quantum gravity~\cite{Sen:2007qy}, apart from exploring the physics of strong gravitational phenomena~\cite{LIGO2017,EventHorizonTelescope:2019dse}.  
Considering for the moment, general thermodynamic systems, a convenient approach to study phase transitions is via a mean field approximation.  Assuming the order parameter to be small and uniform at the transition point, Landau's theory can be applied by making a series expansion of free energy in terms of the small order parameter~\cite{cha95}. Though, this procedure gives useful information for second order phase transitions, applicability of this method for first order transitions is ambiguous, as the order parameter may be large and jump discontinuously. In this situation, a more reliable and unified approach to phase transitions is the one due to Bragg-Williams~\cite{bw1, bw2}, with the possibility of several applications starting from order-disorder transitions in alloys to black holes~\cite{cha95,kubo,Banerjee:2010ve}. The key feature of this approach is to construct a putative off-shell free energy in terms of an appropriate order parameter, whose equilibrium value minimizes the free energy, giving the required phase structure.  While the nature of critical points of general thermodynamic and statistical systems, including black holes, have been studied extensively, it is important to explore neoteric methods for distinctive perspectives.\\

\noindent
Recently,  a certain topological approach\footnote{see~\cite{Cunha:2017qtt,Cunha:2020azh,Guo:2020qwk,Wei:2020rbh,Junior:2021svb} for other motivations following from the study of light-rings} to classify second order critical points has emerged~\cite{Wei:2021vdx} in the context of (extended) thermodynamics of black holes~\cite{Chamblin:1999tk,Caldarelli:1999xj,Kastor:2009wy,Cvetic:2010jb,Dolan:2011xt,Karch:2015rpa,Kubiznak:2012wp,Kubiznak:2016qmn}. The key idea is to start from temperature of the black hole written as a function of entropy $S$ and pressure $P$ and other variables, and then find a potential by eliminating the pressure, such that the zeroes of vector field constructed from this potential correspond to the critical points. Existence of a conserved current ensures that a topological charge can be assigned to each critical point, which can then be used to classify their nature~\cite{Duan:1984ws,Duan:2018rbd,Wei:2021vdx}. In case of systems with multiple critical points or points where new phases appear/disappear a more careful analysis may be required to decipher the true nature of critical points~\cite{Yerra:2022alz}. All the recent works have explored the case of second order critical points in black hole systems~\cite{Wei:2021vdx,Yerra:2022alz,Ahmed:2022kyv,Yerra:2022eov,Wei:2022dzw}, with the phase transition happening in space-time.   Also, the topological classification of critical points in black holes~\cite{Wei:2021vdx,Yerra:2022alz,Ahmed:2022kyv,Yerra:2022eov,Wei:2022dzw} relied heavily on the extended thermodynamic phase space set up, where the cosmological constant is taken to be dynamical, giving rise to  pressure~\cite{Caldarelli:1999xj,Kastor:2009wy,Cvetic:2010jb,Dolan:2011xt,Karch:2015rpa,Kubiznak:2012wp,Kubiznak:2016qmn}. Associating topological nature to critical points should be valid more generally, irrespective of whether the transition happens in real space or in some parameter space, independent of the formalism used to study it. For instance, in the Ising model the order parameter for paramagnetic to ferromagnetic second order phase transition is the magnetic moment. Also, phase transitions in general field theories, such as gauge theories dual to gravity in the bulk via AdS/CFT correspondence, the order parameter is typically some parameter, such as, charge, angular momentum etc.. Further, it was known long back that charged black holes in AdS undergo various phase transitions in the non-extended set up~\cite{Chamblin:1999tk}, though the panoply of such transitions is much richer in extended thermodynamics~\cite{Kubiznak:2016qmn}. It is thus imperative to check whether the topological classification of critical points studied in~\cite{Wei:2021vdx,Yerra:2022alz,Ahmed:2022kyv,Yerra:2022eov,Wei:2022dzw} holds in general thermodynamic situations, such as for first order phase transitions.\\

\noindent 
The aim of this note is to report some progress in addressing the issues posed above.
First, we extend the ideas in~\cite{Wei:2021vdx,Yerra:2022alz,Ahmed:2022kyv,Yerra:2022eov} to show that it is possible to assign topological charge to first order phase transitions as well. We specifically compute the topological charge of the Hawking-Page transition point of black holes in AdS spacetimes.  As we elaborate in the next section, an off-shell formalism such as the Bragg-Williams approach is quite useful in setting up the discussion around the HP transition point. Hawking-Page transition~\cite{Hawking:1982dh} of course continues to evoke remarkable interest, partly due to a dual interpretation in terms of gauge/gravity duality~\cite{,Maldacena:1997re,Gubser:1998bc,Witten:1998qj}, where it corresponds to confinement-deconfinement transition. To check the validity of the result found in the bulk, we also compute the topological charge of the confinement-deconfinement transition from an effective potential approach in dual gauge theory. The two topological charges agree and turn out to be $+1$, even though the order parameters are quite different, i.e., horizon radius $r_+$ for the HP transition and a charge parameter $Q$ for the confinement-deconfinement transition in the boundary gauge theory. Since, the free energy constructed from the Bragg-Williams is off-shell by nature, its saddle points typically give all the equilibrium phases of the system~\cite{chaikin_lubensky_1995,Banerjee:2010ng,Banerjee:2010ve,nayak2008bragg,Dey:2007vt,Dey:2006ds}. Some of the phases are stable and others unstable. By a slight modification of the vector field motivated from~\cite{Wei:2021vdx}, it is possible to assign topological charges to these phases, which correspond to different black hole solutions. The topological charges of stable and unstable black holes turn out to be opposite in character, in agreement with a recent construction explored in a slightly different set up~\cite{Wei:2022dzw}. \\

\noindent
Rest of the note is organized as follows. In section-(\ref{BW}), we explain the basic construction required to understand the Bragg-Williams approach to describe phase transitions of black holes, though the discussion is valid for any general thermodynamic system. Section-(\ref{topology4}) is devoted to the calculation of topological charge of Hawking-Page transition in Schwarzschild and charged black holes in AdS, giving a value $+1$. This computation is then also computed  in the gauge theory via an effective potential constructed using AdS/CFT relations in section-(\ref{boundary}).  In section-(\ref{tcbh}), we compute the topological charges for the equilibrium phases following from the saddle points of free energy. We end summary of our results and some remarks in the concluding section-(\ref{conclusions}).

\section{Hawking-Page  transitions using Bragg-Williams  approach} \label{BW}

\begin{figure}[h!]
	
	{\centering
		
		\subfloat[]{\includegraphics[width=3in]{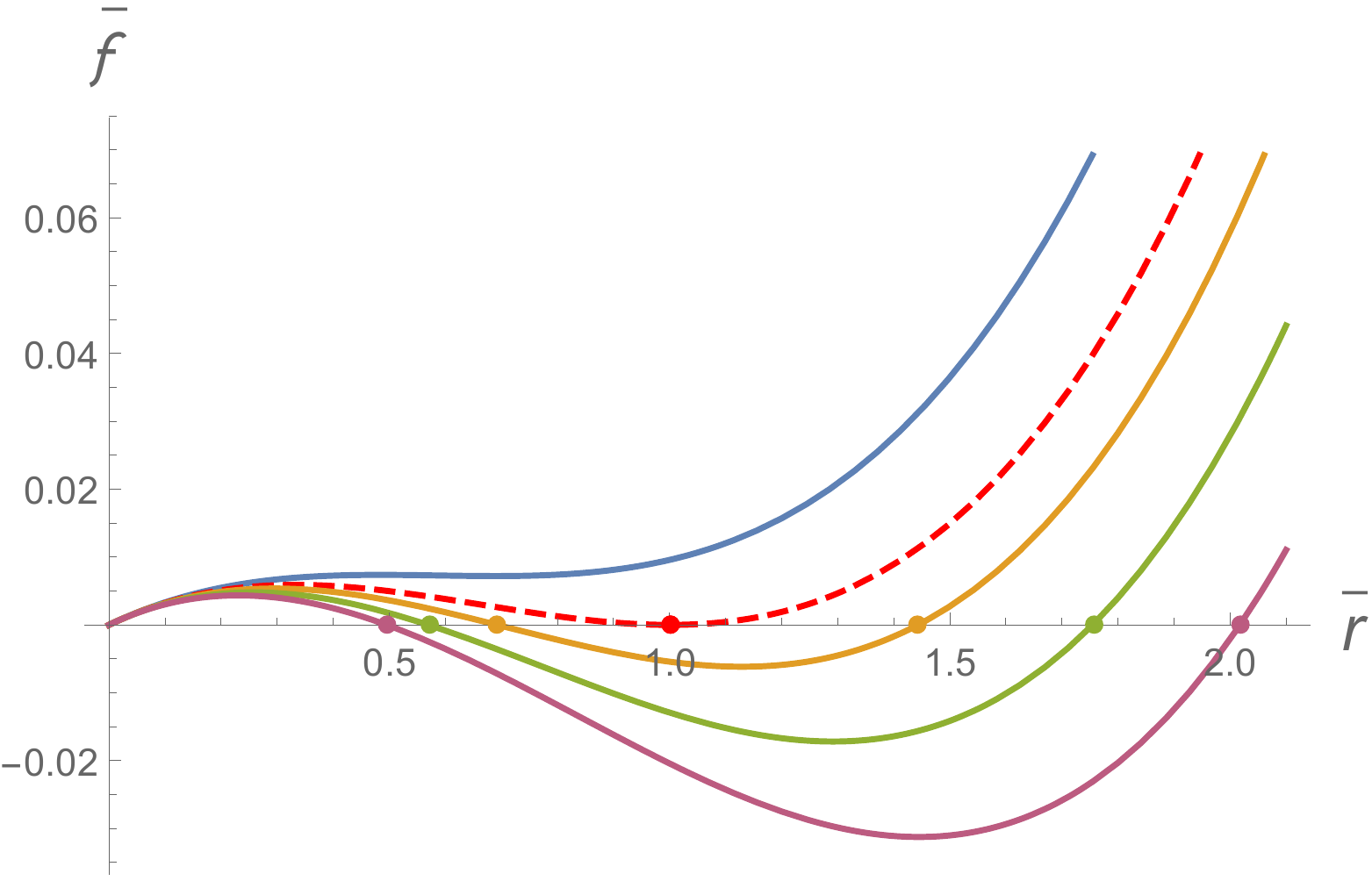}\label{fig:hp_sch_freeplot}}\hspace{0.5cm}	
		\subfloat[]{\includegraphics[width=2.8in]{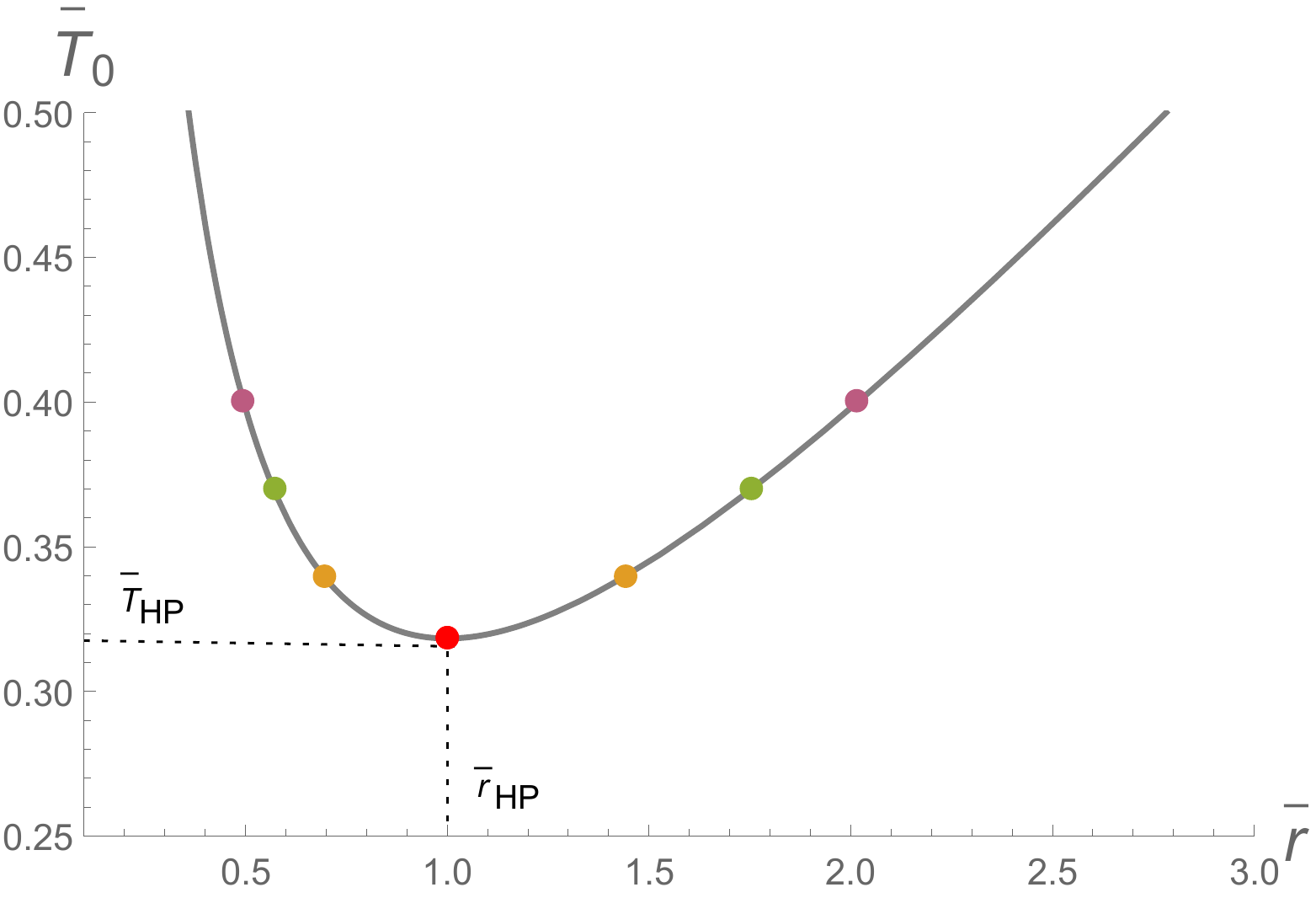}\label{fig:hp_sch_tplot}}	
		
	\caption{\footnotesize For Schwarzschild-AdS black holes:  (a) Behavior of the Bragg-Williams free energy $\bar{f}$, as a function of the order parameter $\bar{r}$ at different temperatures $\bar{T}$. Temperature of the curves increases from top to bottom. Hawking-Page transition happens at the temperature $\bar{T}_{\rm HP}$ (dashed red curve). For $ \bar{T} < \bar{T}_{\rm HP}$, black holes are globally unstable, while for $ \bar{T} > \bar{T}_{\rm HP}$, they are globally stable. (b) The coexistence curve $\bar{T}_{\rm 0}$ of the black hole phase and AdS phase, as a function of $\bar{r}$, shows the HP transition point at its minima (red dot). Other colored dots correspond to respective colored curves in the left plot. Plots are displayed for $n=2$.}
			}
	
\end{figure}

We first present how the Bragg-Williams (BW) method clearly captures the Hawking-Page (HP) transitions for black holes in AdS. The idea is to construct an off-shell free energy function directly from the action, in terms of a suitably chosen order parameter and study the behavior around its saddle points~\cite{chaikin_lubensky_1995,Banerjee:2010ng,Banerjee:2010ve,nayak2008bragg,Dey:2007vt,Dey:2006ds}.
\vskip0.3cm \noindent
Considering the horizon radius $r_+$ as an order parameter, and using the thermodynamic quantities of the black hole, the BW free energy $\bar{f}(\bar{r},\bar{T})$, in $(n+2)$-dimensional spacetime,  can be written as~\cite{Banerjee:2010ve,nayak2008bragg} 
\begin{equation}\label{eq: BWf_sch }
\bar{f}(\bar{r}, \bar{T}) = \bar{E}-\bar{T}\bar{S} = \frac{n\Big( \bar{r}^{n+1} + \bar{r}^{n-1} \Big)}{16\pi}-\bar{T}\frac{\bar{r}^n}{4}. 
\end{equation}
\noindent
Here, the temperature  $\bar{T}=lT$ is a scaled free parameter.  Other quantities, such as the horizon radius $\bar{r}l=r_+$, energy $\bar{E}=lE$, and the entropy $\bar{S}l^n=S$,  are all scaled to absorb the dependence on the AdS length $l$. The function $\bar{f}(\bar{r}, \bar{T})$ along with $\bar{T}$ and $\bar r$ take in general non-equilibrium values. It is only at that minima of the function that all these quantities acquire their equilibrium forms. The behavior of the free energy $\bar{f}(\bar{r}, \bar{T})$, as a function of order parameter $\bar{r}$, for various temperatures is as shown in Fig.~\ref{fig:hp_sch_freeplot}.
\noindent
The AdS phase $(\bar{r}=0)$ is identified with the zero point of free energy. There exists a temperature $\bar{T}_{\rm HP}^{\phantom{HP}}$, below which the black hole free energy is higher than the AdS free energy, thus the black hole phase is globally unstable. However, for temperatures higher than $\bar{T}_{\rm HP}^{\phantom{HP}}$ and thus the black hole phase is globally stable as it has lower free energy than the AdS phase.
Thus, at the temperature $\bar{T}_{\rm HP}^{\phantom{HP}}$, there is a phase transition in which the preferred phase switches from AdS to black holes. This is a first order phase transition as the order parameter $\bar{r}$ shows the discontinuous change,  called as the Hawking-Page (HP) transition. The transition point can be obtained on satisfying the following two conditions simultaneously:
\begin{equation}\label{eq:hp_condition}
\bar{f} = 0, \, \text{and}, \, \frac{\partial \bar{f}}{\partial \bar{r}} = 0.
\end{equation}
\noindent
This gives,
\begin{equation} \label{eq:condition1}
\bar{r}_{\rm HP}^{\phantom{HP}} = 1, \, \text{and}, \, \bar{T}_{\rm HP}^{\phantom{HP}} = \frac{n}{2\pi}.
\end{equation}
\noindent
Alternatively (which will be useful for later purpose), one can also find the Hawking-Page transition point using the first condition of eqn.~\ref{eq:hp_condition}, which gives, 
\begin{equation}\label{eq: t0_sch}
\bar{f}=0 \implies \bar{T}_{\rm 0}^{\phantom{0}} = \frac{n}{4\pi}\Big(\bar{r} + \frac{1}{\bar{r}} \Big).
\end{equation}
\noindent
The Hawking-Page transition point can then be located at the  minima of the curve $\bar{T}_{\rm 0}^{\phantom{0}}$, as shown in the Fig.~\ref{fig:hp_sch_tplot}.
Next, we consider the case of Hawking-Page transition exhibited by charged-AdS black holes in the grand canonical ensemble (i.e., fixed potential $\bar{\mu}$).
The corresponding Bragg-Williams free energy function $\bar{f}(\bar{r}, \bar{T},\bar{\mu})$, turns out to be~\cite{Banerjee:2010ve,Banerjee:2010ng}:
\begin{equation} \label{eq:bwf}
\bar{f}(\bar{r}, \bar{T},\bar{\mu}) = \bar{E}-\bar{T}\bar{S}-\bar{\mu}\bar{Q}=n\bar{r}^{n-1}(1-c^2\bar{\mu}^2)-4\pi\bar{r}^n\bar{T}+n\bar{r}^{n+1}.
\end{equation}
\noindent
Here, $\bar{T}$ and $\bar{\mu}$ are treated as the external parameters, $\bar{Q}$ is the charge of the black hole and $c=\sqrt{2(n-1)/n}$.
In this case, the HP transition happens at (from the eqn~\ref{eq:hp_condition})
\begin{equation}\label{eq:rhp and thp_RN}
\bar{r}_{\rm HP}^{\phantom{HP}} = \sqrt{1-c^2\bar{\mu}^2}, \, \text{and}, \, \bar{T}_{\rm HP}^{\phantom{HP}} = \frac{n\sqrt{1-c^2\bar{\mu}^2}}{2\pi}.
\end{equation}
\noindent
Further, the coexistence curve $\bar{T}_{\rm 0}^{\phantom{0}} (\bar{r},\bar{\mu})$ of the black hole phase and AdS phase becomes (from the condition $\bar{f} =0$),
\begin{equation}\label{eq:to_RN}
\bar{T}_{\rm 0}^{\phantom{0}} (\bar{r},\bar{\mu}) = \frac{n}{4\pi}\Big( \bar{r} + \frac{(1-c^2 \bar{\mu}^2)}{\bar{r}}\Big).
\end{equation}
\noindent
One can see that, for a fixed potential $\bar{\mu}$, the behaviors of the free energy $\bar{f}$ and the coexistence curve $\bar{T_{\rm 0}^{\phantom{0}}}$, are similar to the previous case.  
In the following section, we make a set up to find the topological charge associated with the Hawking-Page transition point in Schwarzschild-AdS black hole case and charged-AdS black hole case as well.
\section{Assigning Topological charge to Hawking-Page transition}\label{topology4}

\begin{figure}[h!]
	{\centering
		\subfloat[]{\includegraphics[width=3in]{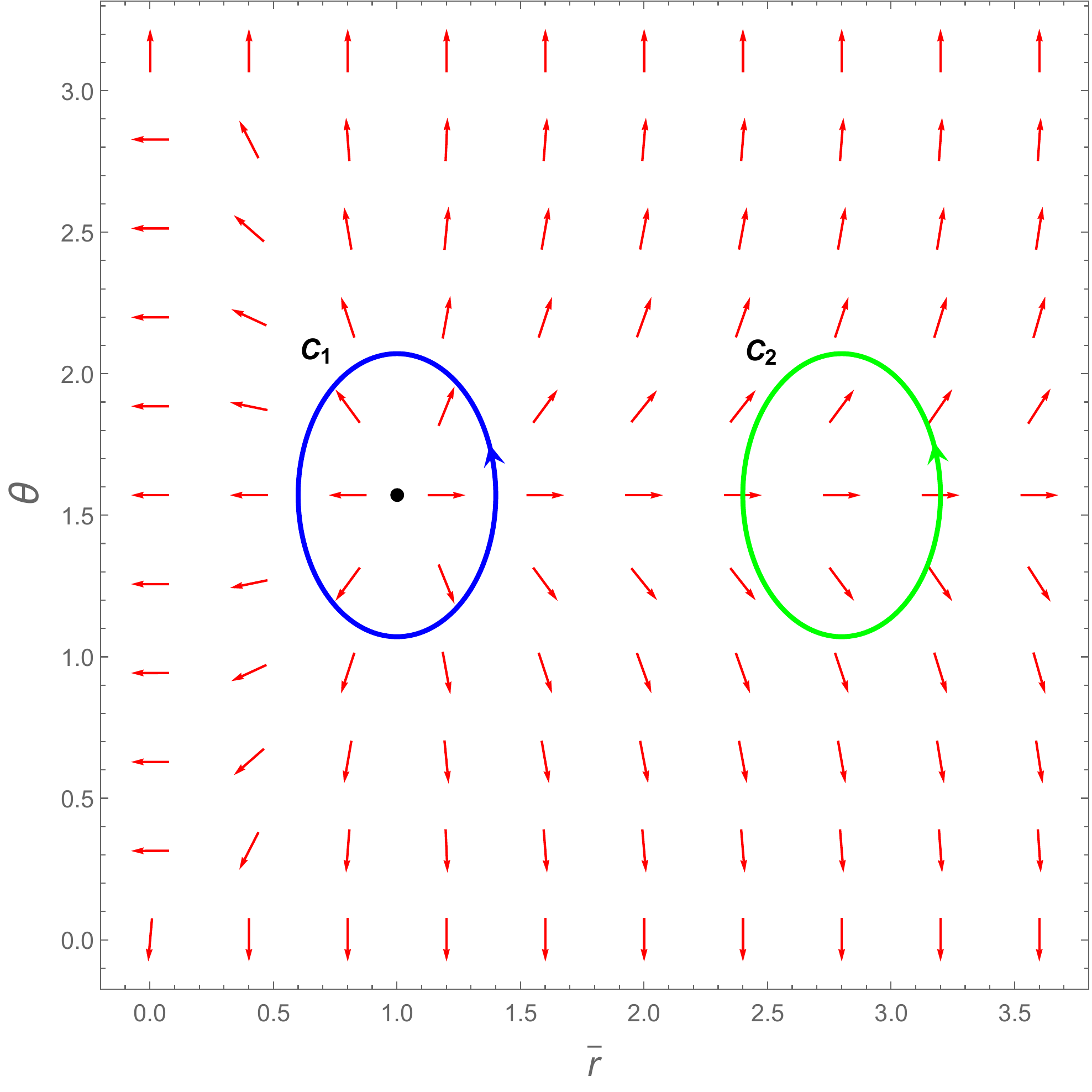}\label{Fig:hp_sch_vecplot}}\hspace{0.5cm}	
		\subfloat[]{\includegraphics[width=3in]{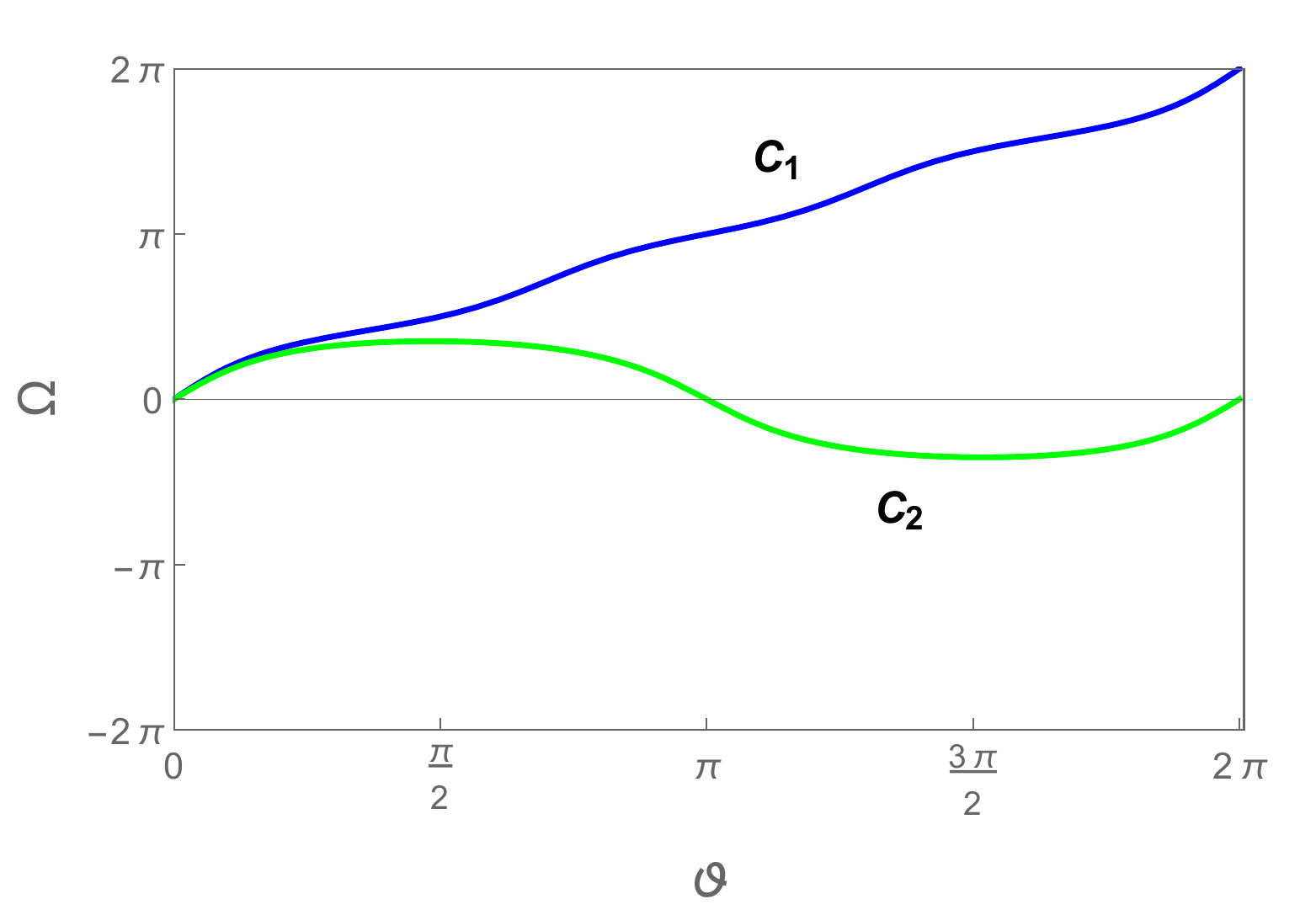}\label{Fig:hp_sch_omegaplot}}				
		
		\caption{\footnotesize For Schwarzschild-AdS black holes: (a) The  vector field $n$ in the $\theta-\bar{r}$ plane, shows the presence of Hawking-Page (HP) transition point (black dot, located at $\bar{r}_{\rm HP} = 1$) at $\phi =0$. Contour $C_1$ contains the HP transition point, while the contour $C_2$ does not. 
			 (b) $\Omega$ vs $\vartheta$ for contours $C_1$ (blue curve),   and $C_2$ (green curve). Plots are displayed for $n=2$, and the parametric coefficients of the contours are $(a,b,r_0) =(0.4,0.5,1)$ for  $C_1$, and $(0.4,0.5,2.8)$ for $C_2$.} 
	}
\end{figure}

We first consider the Schwarzschild-AdS black holes case. In order to assign the topological charge to HP transition point, we employ the temperature $\bar{T}_{\rm 0}^{\phantom{0}}$ (eqn.~\ref{eq: t0_sch}) obtained from the Bragg-Williams free energy landscape, which serves to define the vector filed $\phi(\phi^{\bar{r}}, \phi^{\theta})$~\cite{Wei:2021vdx}, in the following way:
\begin{eqnarray}
\phi^{\bar{r}} &=& \partial_{\bar{r}} \Phi = \frac{n}{4\pi \text{sin} \theta}\Big( 1- \frac{1}{\bar{r}^2}\Big), \\
\phi^{\theta} &=& \partial_{\theta} \Phi = -\frac{n \text{cot}\theta \, \text{csc}\theta}{4\pi} \Big( \bar{r}+ \frac{1}{\bar{r}}\Big),
\end{eqnarray}
where, $\Phi=\frac{1}{\text{sin}\theta} \bar{T}_{\rm 0}^{\phantom{0}} (\bar{r})$. A key outcome of this procedure is the existence of a topological current $j^{\mu}$ satisfying the condition $\partial_{\mu}\, j^{\mu}=0$, which is non-zero only at the points where the vector field $\phi^a$ is identically zero, i.e., $\phi^a(x^i) = 0$.  In the present case, this vector field $\phi$ vanishes exactly at the Hawking-Page transition point, which can be seen clearly from the normalized vector field  $n=(\frac{\phi^{\bar{r}}}{||\phi||},\frac{\phi^\theta}{||\phi||})$  plot in the Fig.~\ref{Fig:hp_sch_vecplot}.
The definition of topological charge ensues from the above construction as~\cite{Duan:1984ws,Duan:2018rbd,Wei:2021vdx}
\begin{equation}\label{tcharge}
 Q_t =\int_\Sigma j^0 d^2x= \Sigma_{i=1} w_i\, ,
\end{equation}
contained in a region $\Sigma$. Here, $w_i$ denotes the winding number of the $i$-th point corresponding to zero of $\phi$. If now $\Sigma$ encompasses the available parameter space of the thermodynamic system, the phase transition points of thermodynamic systems can fall into different topological classes.
This can be seen from the fact that, $Q_t$ can be positive or negative, with the possibility of total topological charge being zero as well. To be precise, we consider a piece-wise smooth (positively oriented) contour $C$,  in a certain $\theta - \bar r$ plane,  which we chose to parameterize by the angle $\vartheta \in (0, 2\pi)$, as~\cite{Wei:2021vdx,Wei:2020rbh}:
\begin{eqnarray}
	\left\{
	\begin{aligned}
		\bar r&=a\cos\vartheta+r_0, \\
		\theta&=b\sin\vartheta+\frac{\pi}{2}.
	\end{aligned}
	\right.
\end{eqnarray} \noindent
Along the contour $C$, the deflection angle $\Omega({\vartheta})$ of $\phi$ is,
\begin{equation}
\Omega(\vartheta)=\int_{0}^{\vartheta}\epsilon_{ab}n^{a}\partial_{\vartheta}n^{b}d\vartheta,
\end{equation}
Whose integration reveals the topological charge
\begin{equation}
Q_t = w_i=\frac{1}{2\pi} \Omega (2\pi) \, . \label{eq: qt_from_Omega}
\end{equation} 

\begin{figure}[h!]
	{\centering
		\subfloat[]{\includegraphics[width=3in]{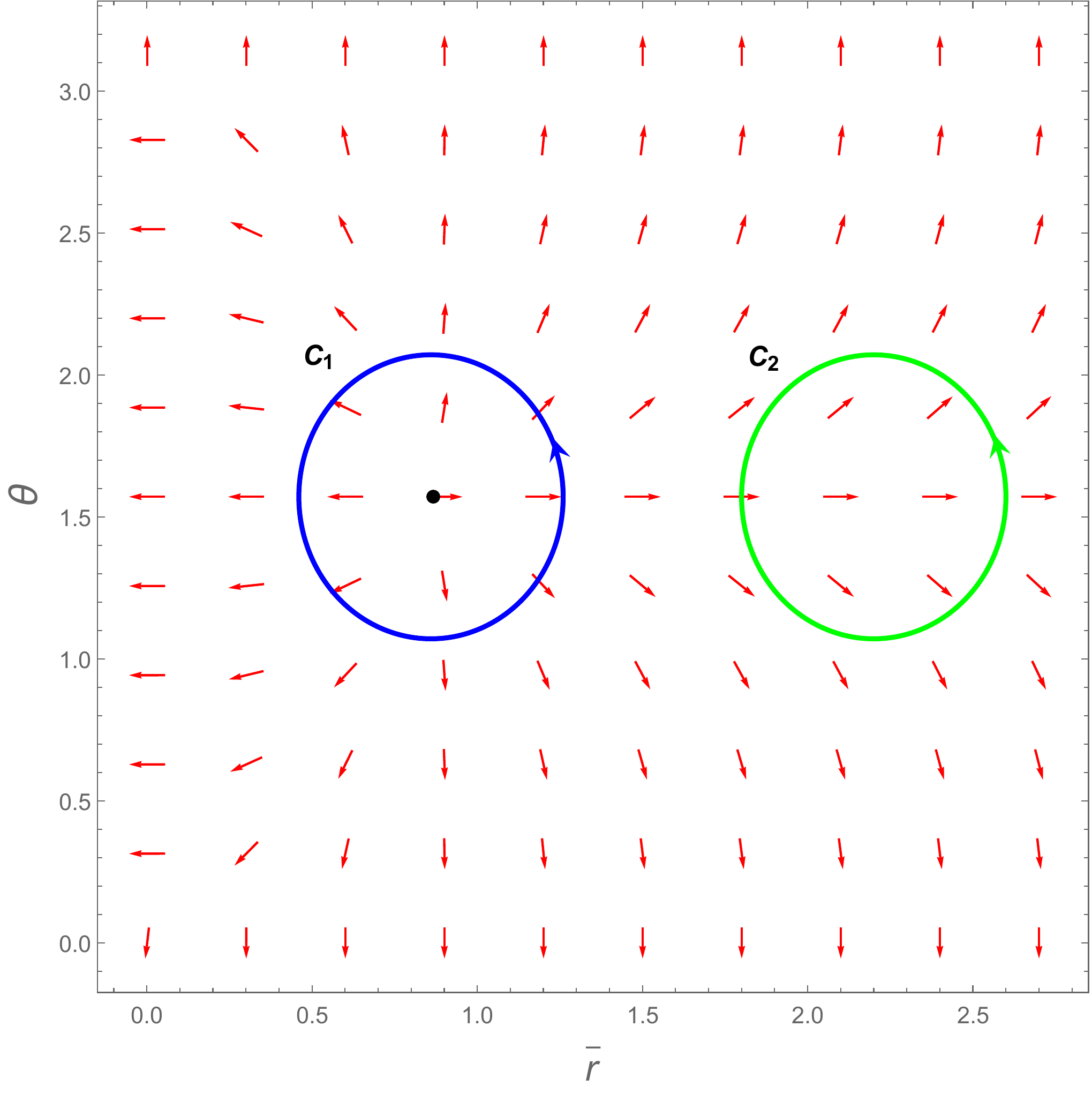}\label{Fig:hp_RN_vecplot}}\hspace{0.5cm}	
		\subfloat[]{\includegraphics[width=3in]{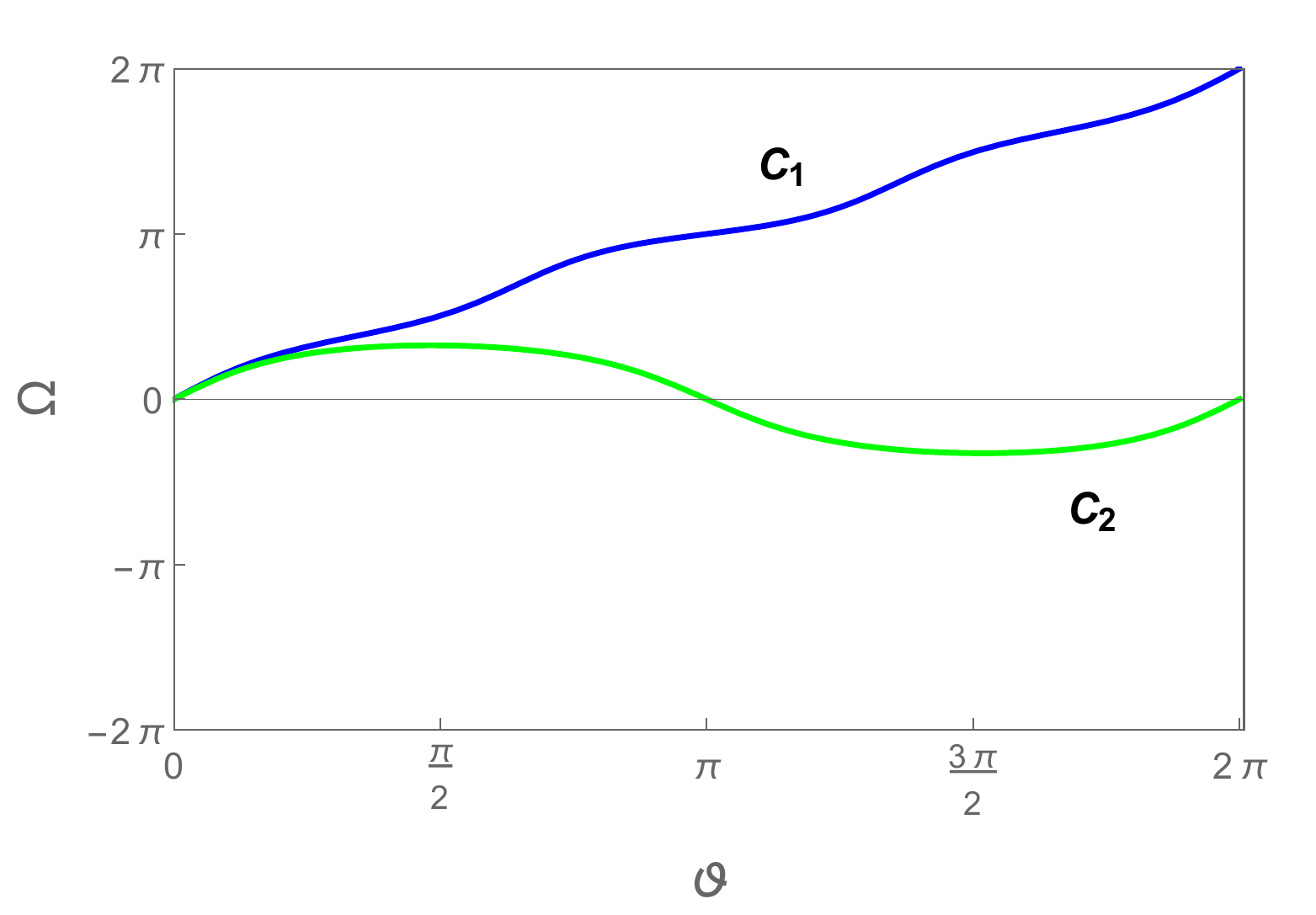}\label{Fig:hp_RN_omegaplot}}				
		
		\caption{\footnotesize For charged-AdS black holes: (a) The  vector field $n$ in the $\theta-\bar{r}$ plane, shows the presence of Hawking-Page (HP) transition point (black dot, located at $\bar{r}_{\rm HP} = \sqrt{1-c^2\bar{\mu}^2}=0.866$) at $\phi =0$. Contour $C_1$ contains the HP transition point, while the contour $C_2$ does not. 
			(b) $\Omega$ vs $\vartheta$ for contours $C_1$ (blue curve),   and $C_2$ (green curve). Plots are displayed for $n=2$, $\bar{\mu}=0.5$, and the parametric coefficients of the contours are $(a,b,r_0) =(0.4,0.5,0.86)$ for  $C_1$, and $(0.4,0.5,2.2)$ for $C_2$.} 
	}
\end{figure}

\noindent
The behavior of the deflection angle $\Omega(\vartheta)$, for the given contours in the Fig.~\ref{Fig:hp_sch_vecplot}, is as shown in Fig~\ref{Fig:hp_sch_omegaplot}, from where we find that the topological charge associated with the HP transition point would be $Q_t\big|_{\rm HP}^{\phantom{HP}} = \frac{1}{2\pi} \Omega (2\pi) = +1$ (given by the contour $C_1$).
\vskip 0.5cm \noindent
We note here that, if one uses the black hole temperature $\bar{T}_{\rm BH}^{\phantom{BH}}$ (obtained from the condition $\frac{\partial \bar{f}}{\partial \bar{r}} =0$, in eqn.~\ref{eq:hp_condition}), instead of  $\bar{T}_{\rm 0}$, to define the vector field $\phi$, then
one could assign the topological charge to the point where the black hole possesses the lowest temperature ($\bar{T}_{min}$)~\footnote{$\bar{T}_{min} = \frac{\sqrt{n^2-1}}{2\pi}$.}.  
In fact the topological charge for this point is $+1$. However, we do not consider this situation, as the black hole at this point is globally unstable.  
\vskip 0.5cm \noindent
Next, we consider the Hawking-Page transition in  charged-AdS black holes case. The vector field $\phi$, in this case, turns out to be (employing eqn.~\ref{eq:to_RN}):
\begin{eqnarray}
\phi^{\bar{r}} &=& (\partial_{\bar{r}} \Phi )_{\theta, \bar{\mu}} = \frac{n (\bar{r}^2+c^2\bar{\mu}^2-1)}{4\pi \bar{r}^2 \text{sin} \theta }, \\
\phi^{\theta} &=& (\partial_{\theta} \Phi)_{\bar{r}, \bar{\mu}} = -\frac{n \text{cot}\theta \, \text{csc}\theta (1+\bar{r}^2-c^2\bar{\mu}^2)}{4\pi \bar{r}},
\end{eqnarray}
where, $\Phi=\frac{1}{\text{sin}\theta} \bar{T}_{\rm 0}^{\phantom{0}} (\bar{r}, \bar{\mu})$. The vanishing of the vector field $\phi$ at the HP transition point can be seen from the Fig.~\ref{Fig:hp_RN_vecplot}.
The behavior of the deflection angle $\Omega(\vartheta)$ along the contours $C_1$ and $C_2$, shown in the Fig.~\ref{Fig:hp_RN_omegaplot}, revels that the associated topological charge with the HP transition point would be again $+1$ (given by the contour $C_1$).

\section{Topological charge of confinement-deconfinement transition}\label{boundary}

In section-(\ref{BW}), an off-shell free energy was employed which captured the stable, unstable and metastable phases of black holes in AdS. A topological charge was assigned specifically to the Hawking-Page transition point in section-(\ref{topology4}). It is also possible to assign topological charge to each of the phases of the system, resulting in a topological classification of black hole solutions in different regions of the thermodynamic phase space. Before doing this, we first indicate how results of section-(\ref{topology4}) can be extended to a boundary field theory set up, where the phase transition may be studied following from an appropriate potential, in terms of an order parameter different from the bulk. Sometime back, it was shown how given a free energy in the bulk, an off-shell phenomenological effective potential can be constructed in the gauge dual, whose equilibrium points exactly correspond to various phase of the theory. We should note that a general construction of an effective potential directly in the gauge theory is a non-trivial task, but the AdS/CFT conjecture allows a slightly different route. Motivated from the free energy in eqn. (\ref{eq:bwf}), an effective potential (which may not be unique) in the gauge theory dual to charged-AdS black holes (in grand canonical ensemble) can be constructed as~\cite{Banerjee:2010ng}:
\begin{figure}[h!]		
\centering {		{\includegraphics[width=3in]{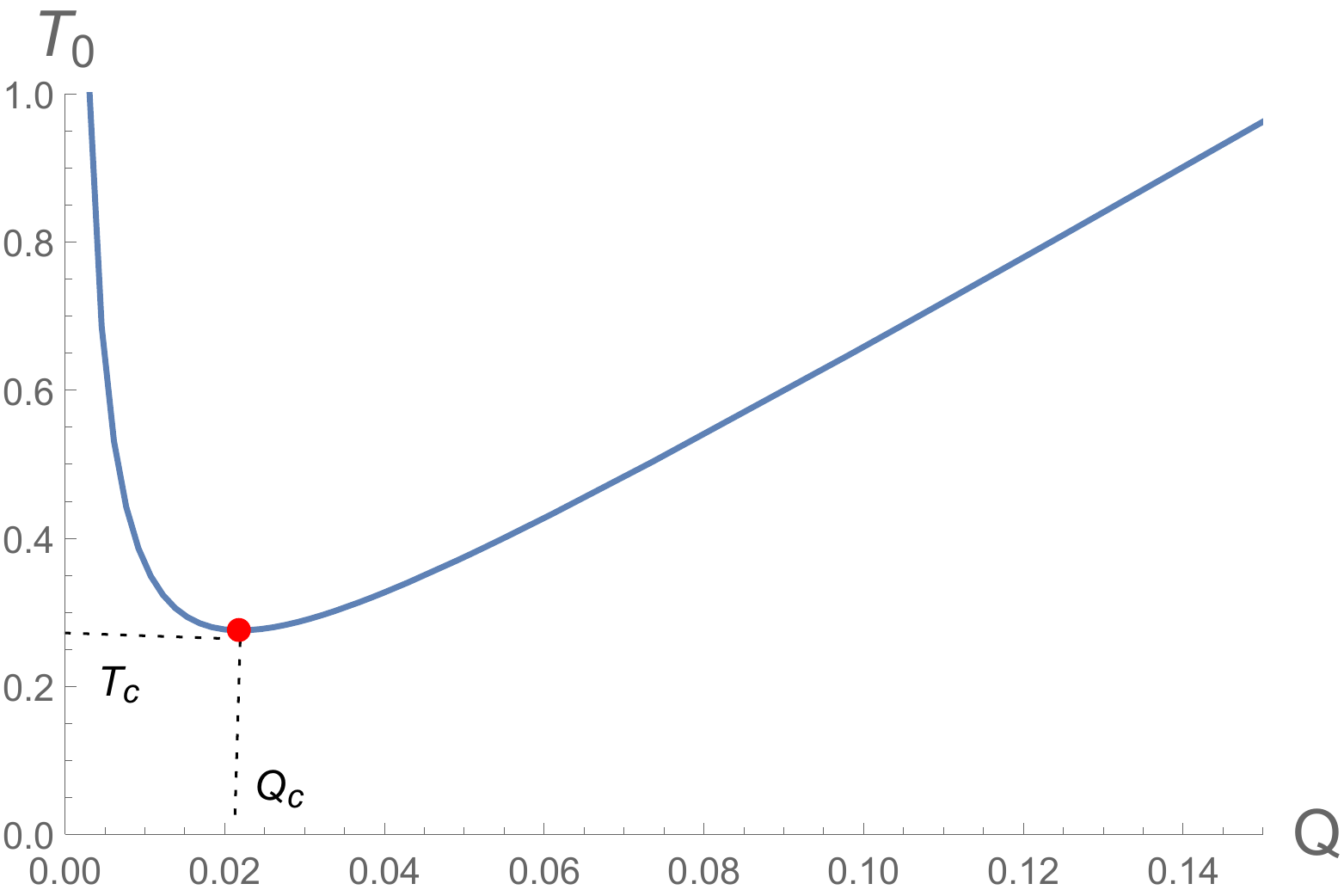}		
		\caption{\footnotesize $T_0$ for the boundary gauge theory, whose minimum gives $T_c$. Here, we set $n=2, l=1, \mu=0.5.$ }
\label{fig:T0_bdg}	}}
\end{figure}
\begin{equation}
W=\frac{\omega_n\,N_c^2}{8 \pi ^2} \left\{ \frac{2 \pi ^2 n \left(1-c^2 \mu ^2\right)  Q}{(n-1) \mu} - T\Big[ \pi(2\pi)^{3n-2} \Big( \frac{Q}{(n-1)\mu}\Big)^n \Big]^{\frac{1}{n-1}} +  \frac{n}{l^2}\Big( \frac{2\pi^2 Q}{(n-1) \mu} \Big)^{\frac{n+1}{n-1}}   \right\},
\end{equation}
 where, $N_c$ stands for number of colors. The critical temperature for the confinement deconfinement transition can be obtained on satisfying the following two conditions simultaneously:
\begin{equation}\label{eq:cd_condition}
W = 0, \, \text{and}, \, \frac{\partial W}{\partial Q} = 0.
\end{equation}
where $Q$ is treated as the order parameter for the transition.
\begin{figure}[h!]
	{\centering
		\subfloat[]{\includegraphics[width=3in]{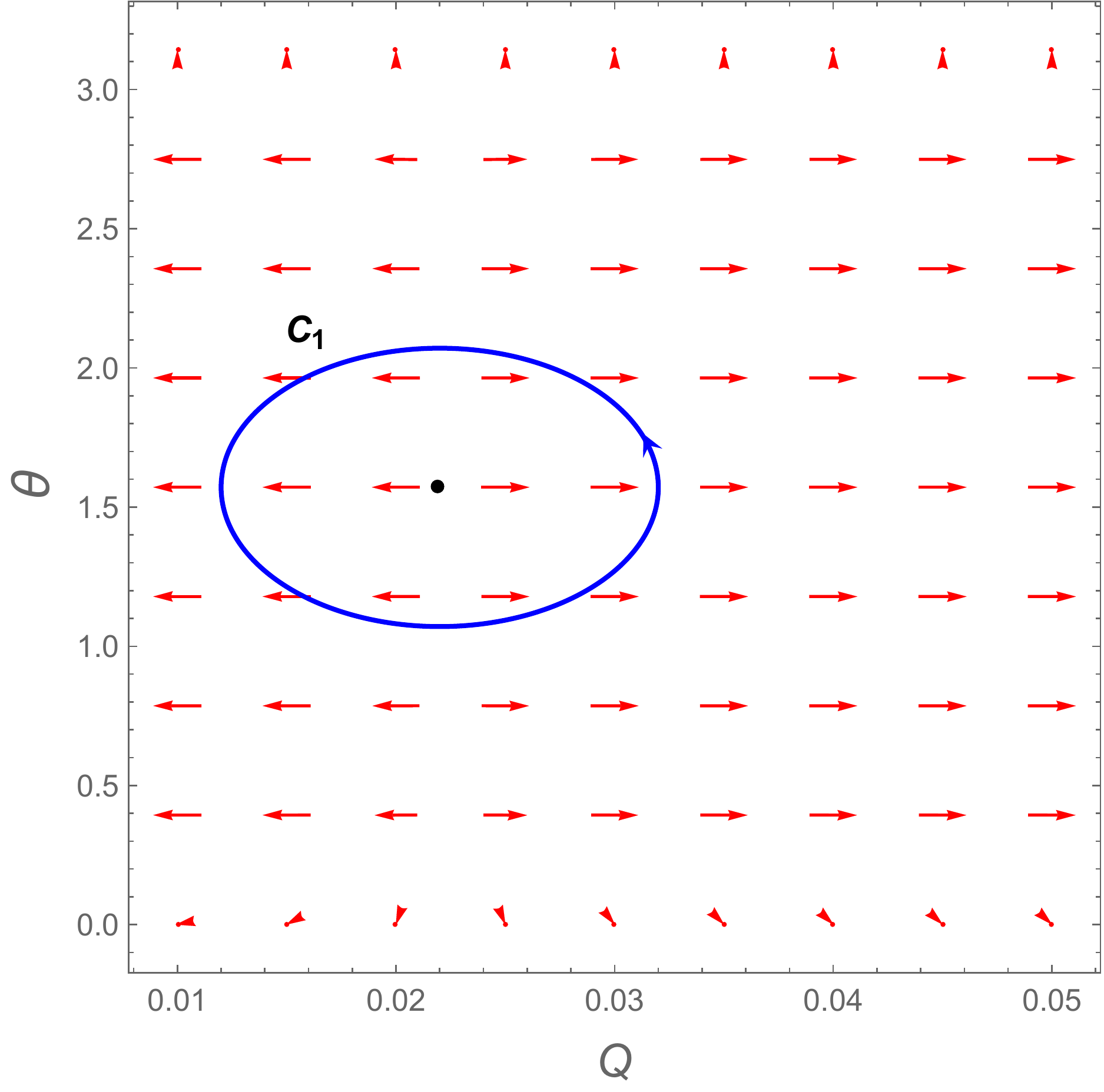}\label{Fig:hp_RN_bdg_vecplot}}\hspace{0.3cm}	
		\subfloat[]{\includegraphics[width=3in]{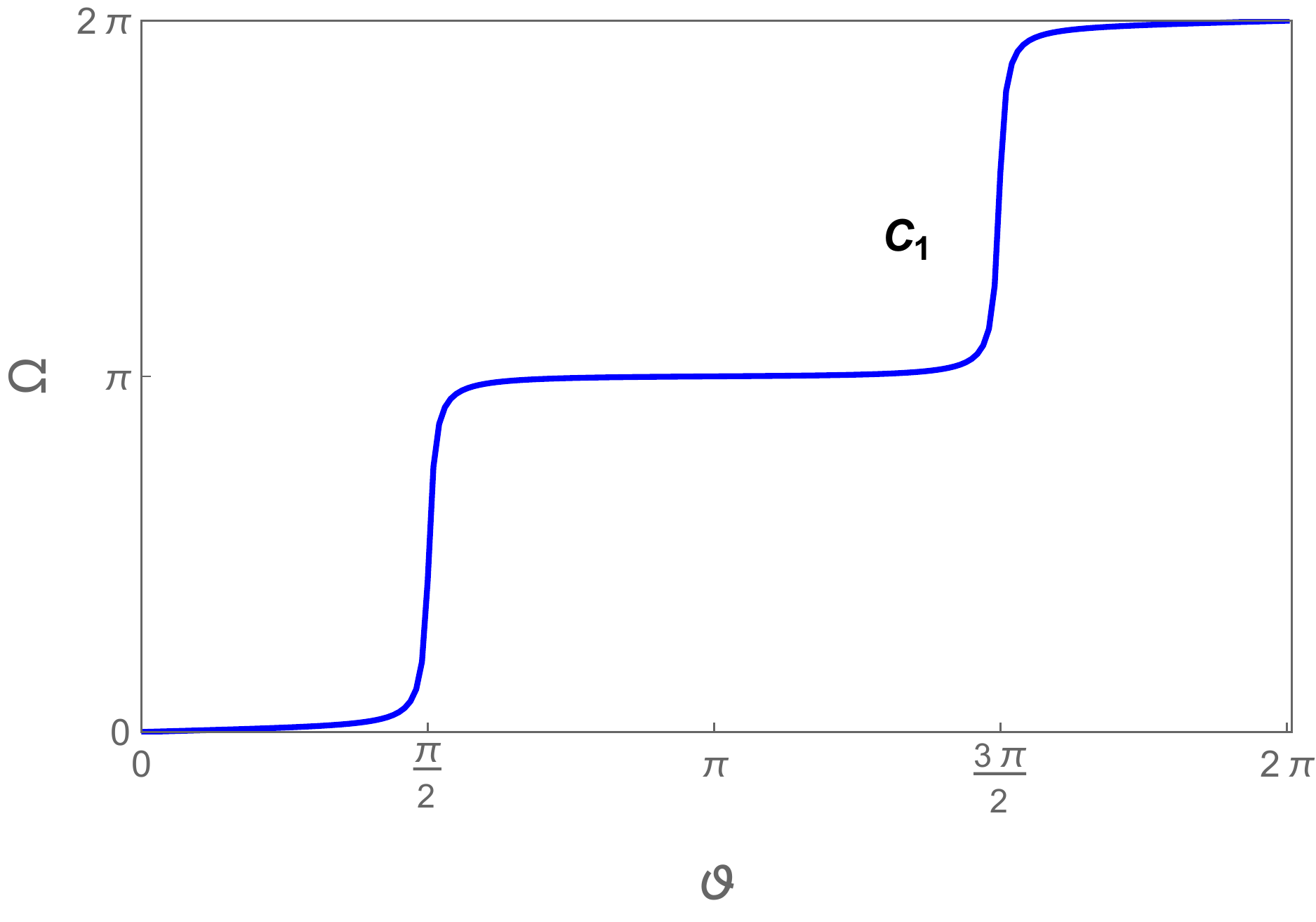}\label{Fig:hp_RN_bdg_omegaplot}}				
		
		\caption{\footnotesize For boundary guage theory: (a) The  vector field $\phi$ vanishes at the confinement-deconfinement transition point (black dot, located at $Q_{\rm c}$). Contour $C_1$ contains the  transition point. 
			(b) $\Omega$ vs $\vartheta$ for contour $C_1$. Plots are displayed for $n=2$, $l=1$, $\mu=0.5$.} 
	}
\end{figure}
\begin{figure}[t]
	{\centering
		\subfloat[]{\includegraphics[width=2in]{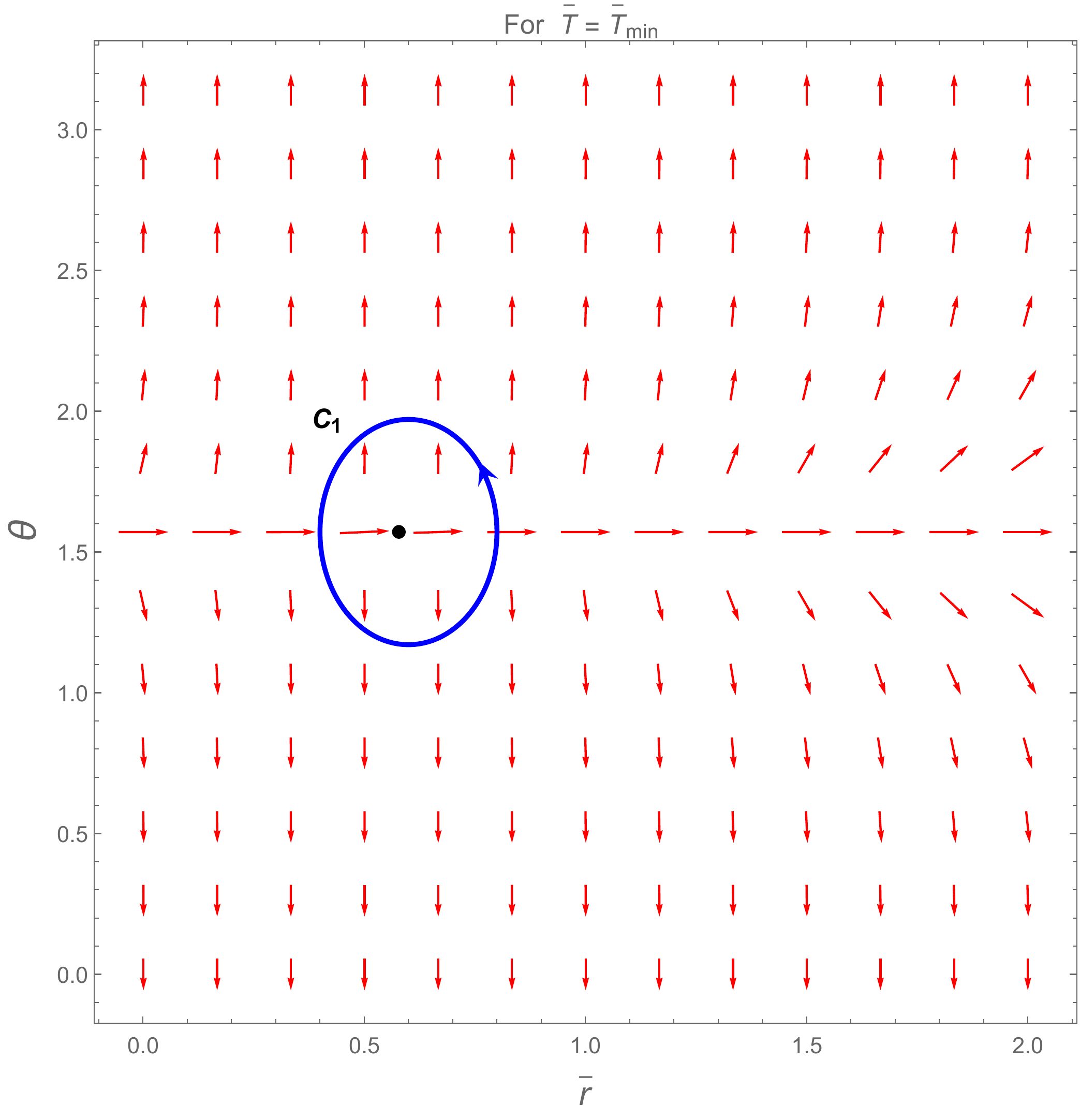}}\hspace{0.3cm}	
		\subfloat[]{\includegraphics[width=2.5in]{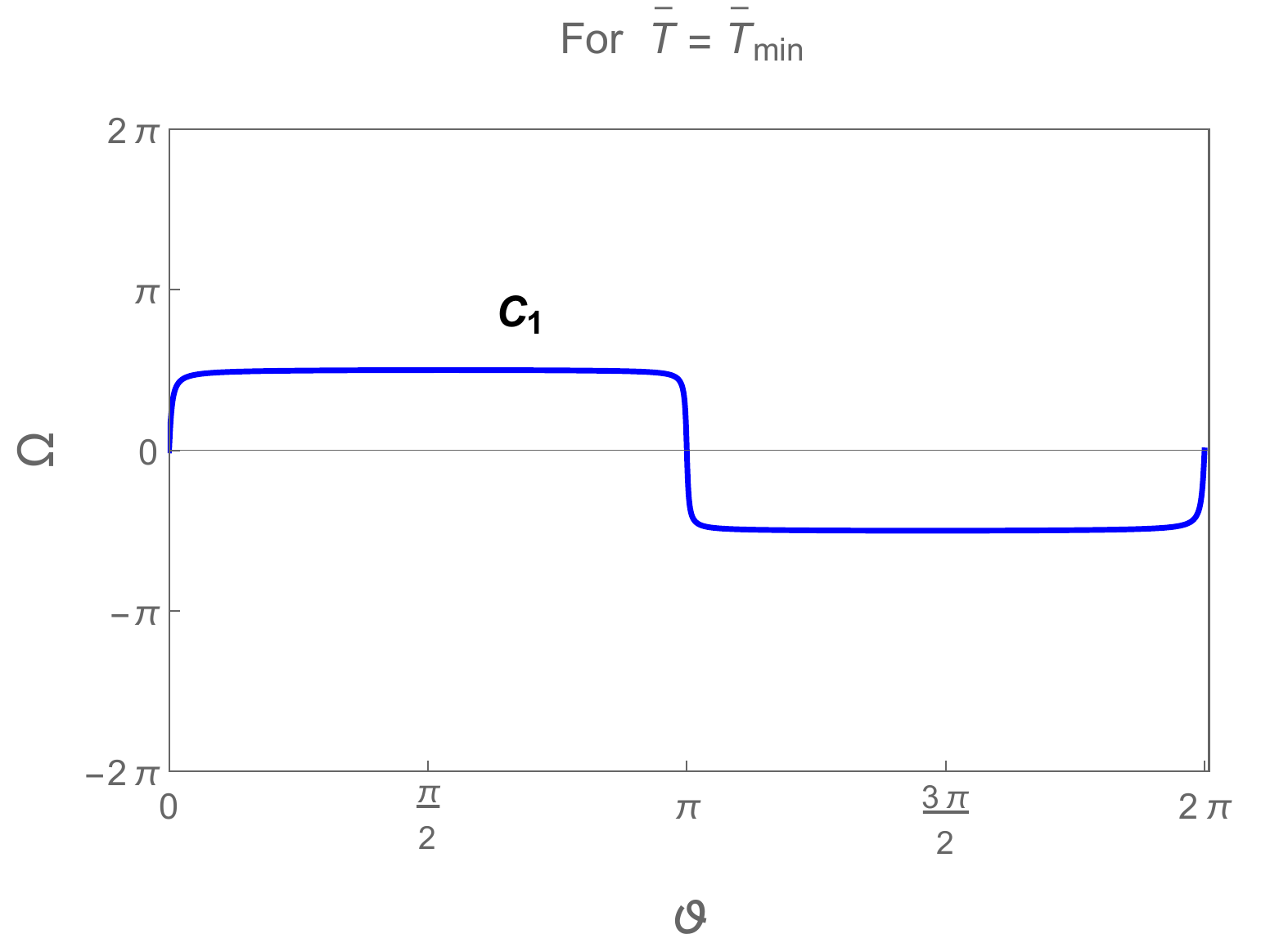}}\hspace{0.3cm}
		\subfloat[]{\includegraphics[width=2in]{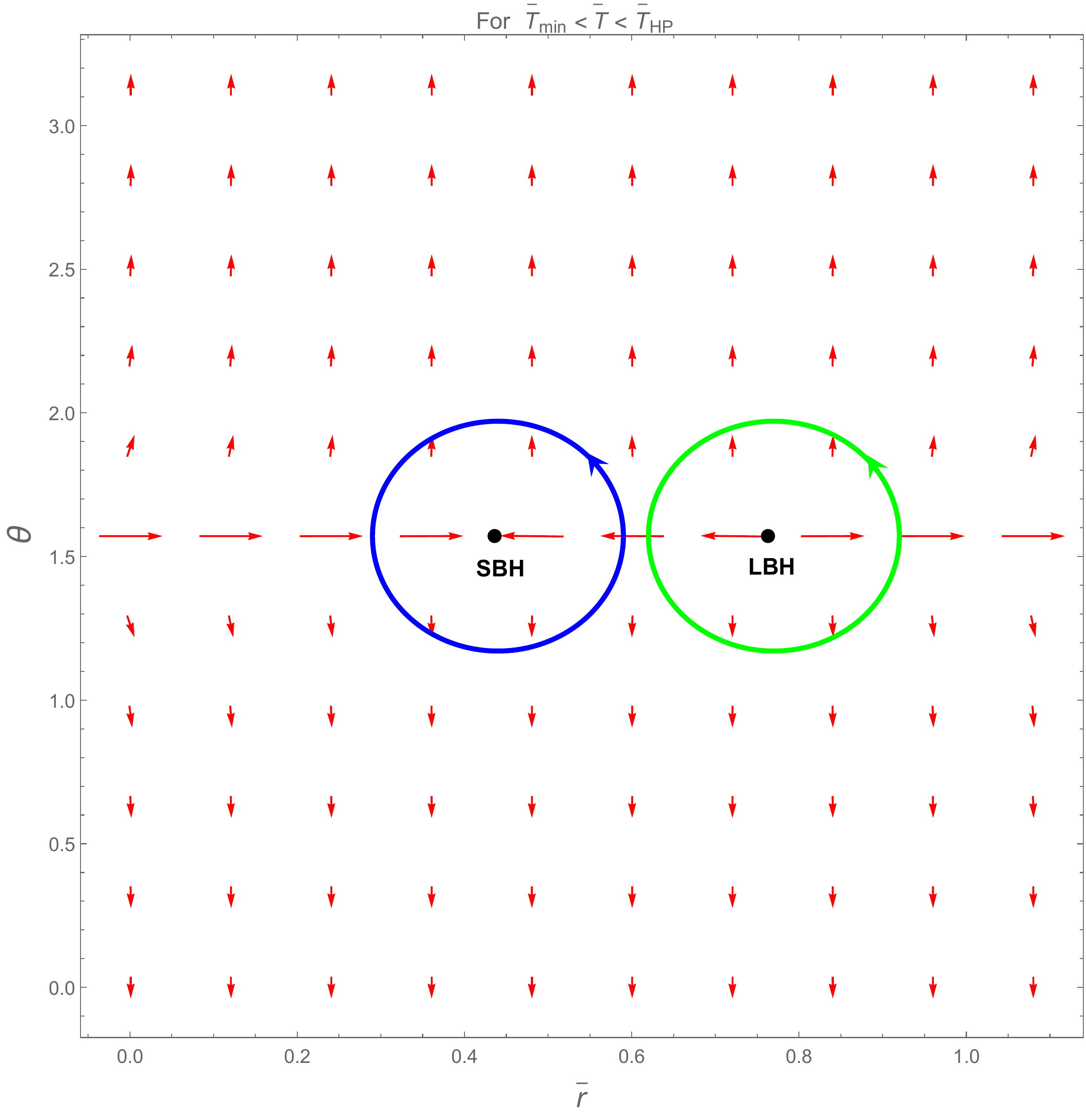}}	\hspace{0.3cm}
		\subfloat[]{\includegraphics[width=2.5in]{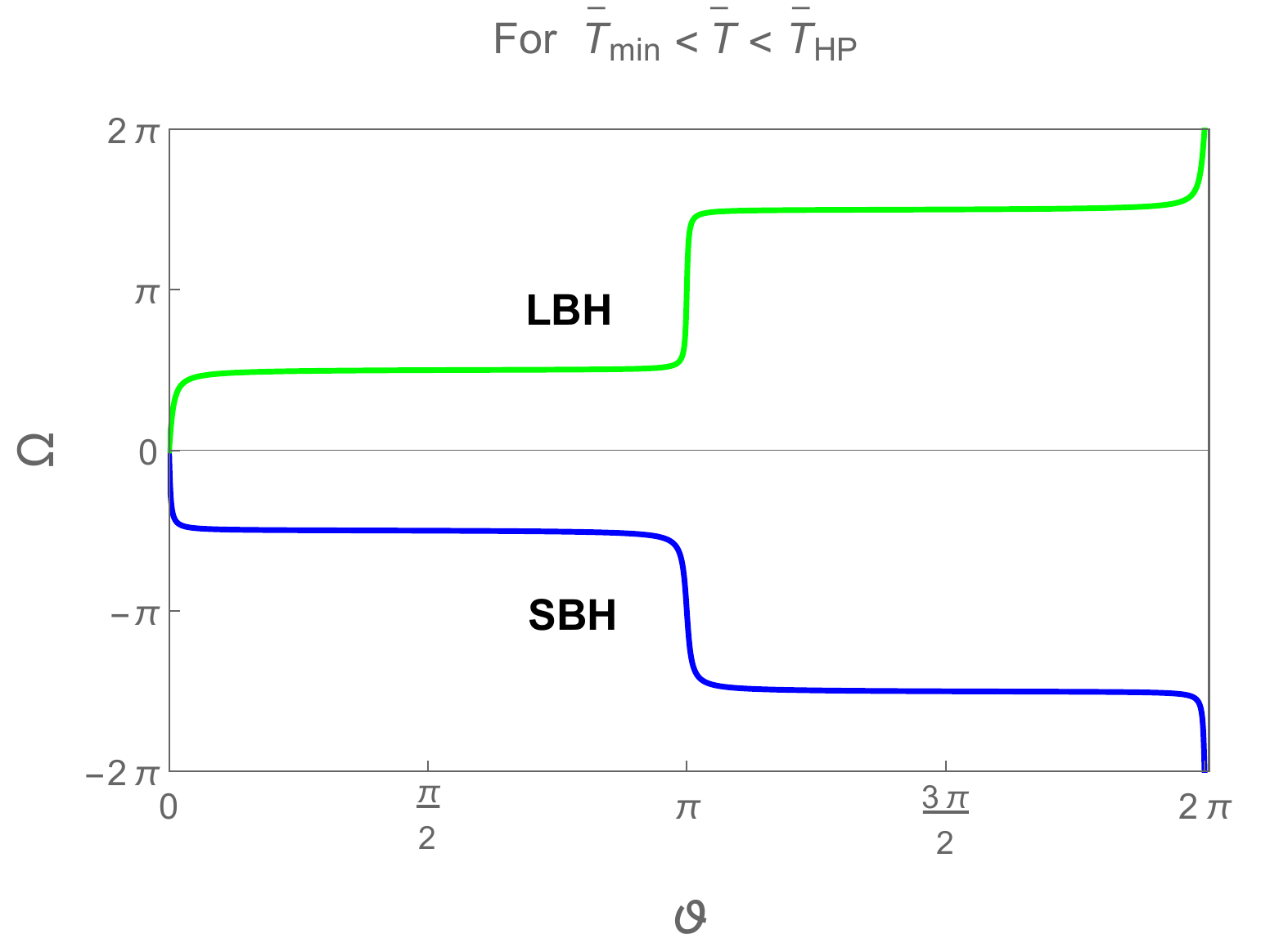}}\hspace{0.3cm}
		\subfloat[]{\includegraphics[width=2in]{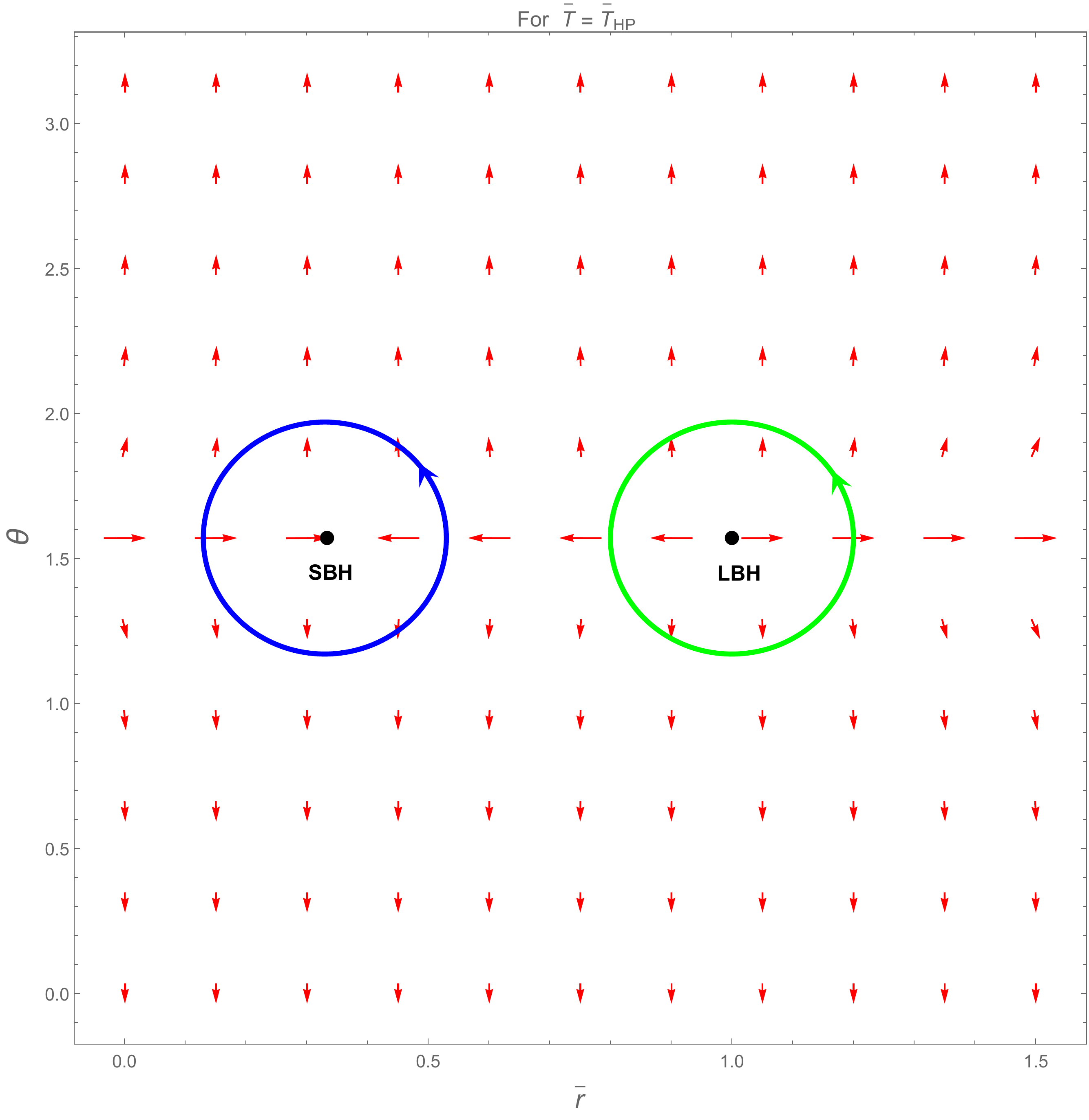}}\hspace{0.3cm}	
		\subfloat[]{\includegraphics[width=2.5in]{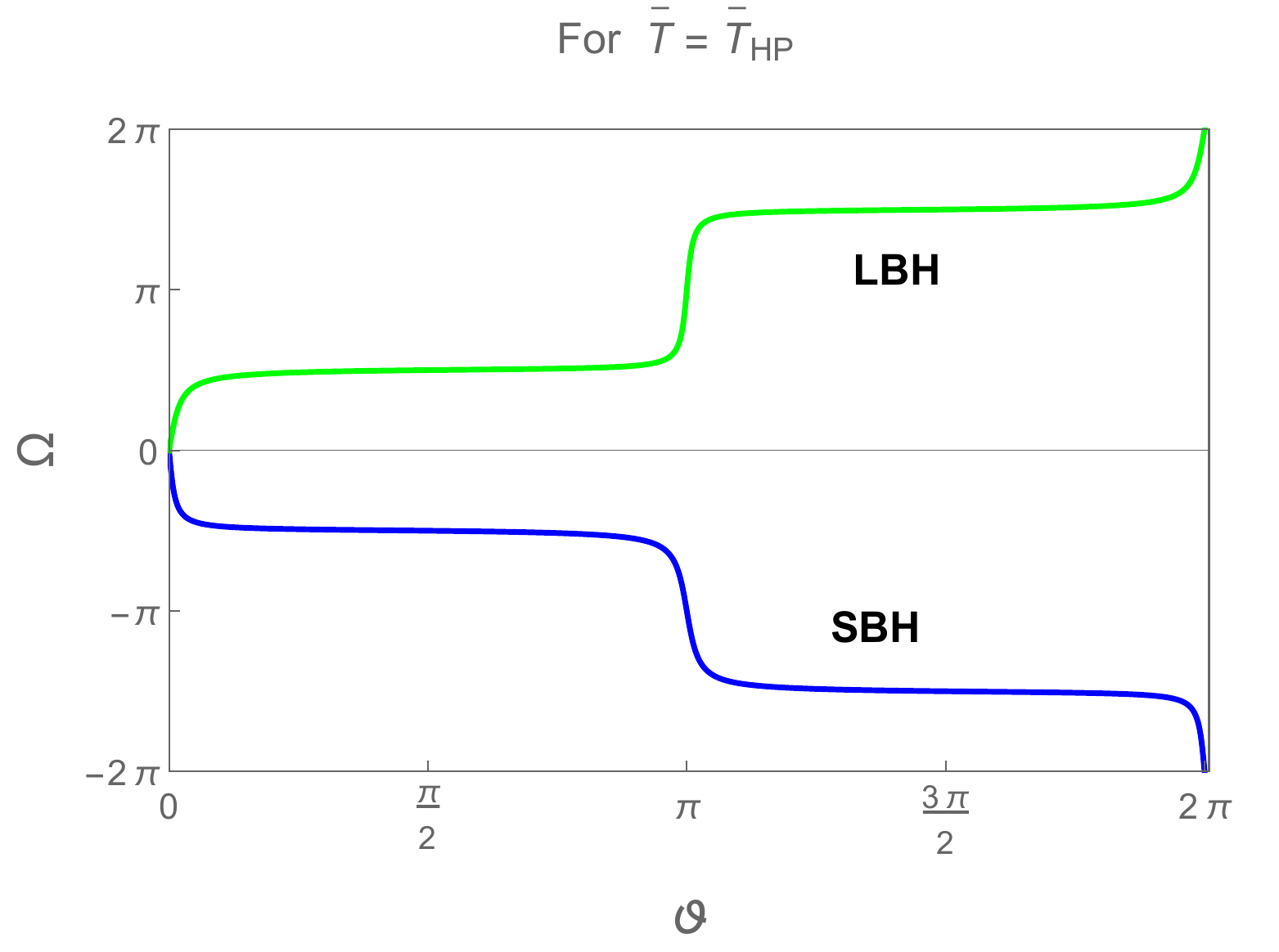}}\hspace{0.3cm}
		\subfloat[]{\includegraphics[width=2in]{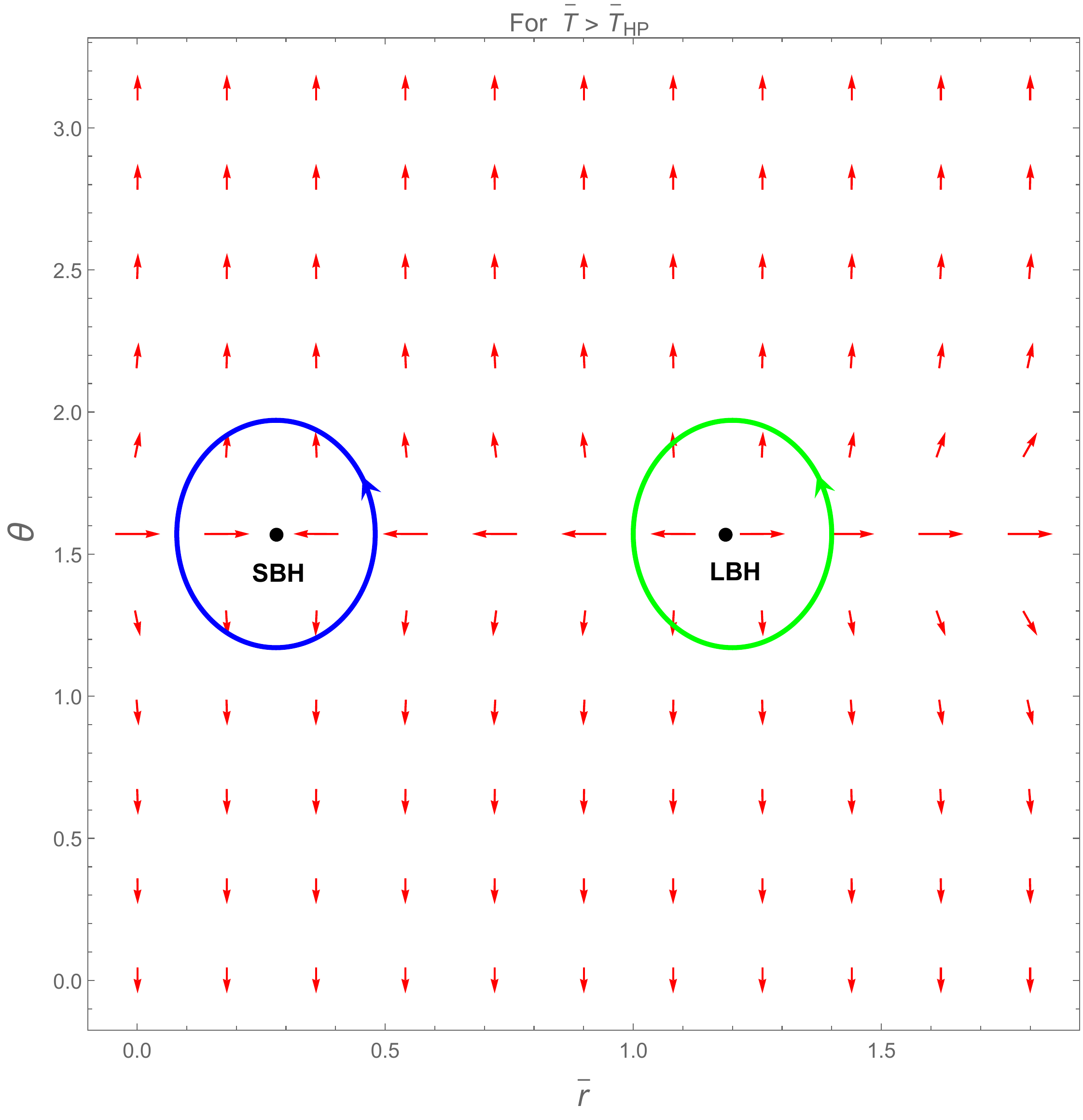}}\hspace{0.3cm}	
		\subfloat[]{\includegraphics[width=2.5in]{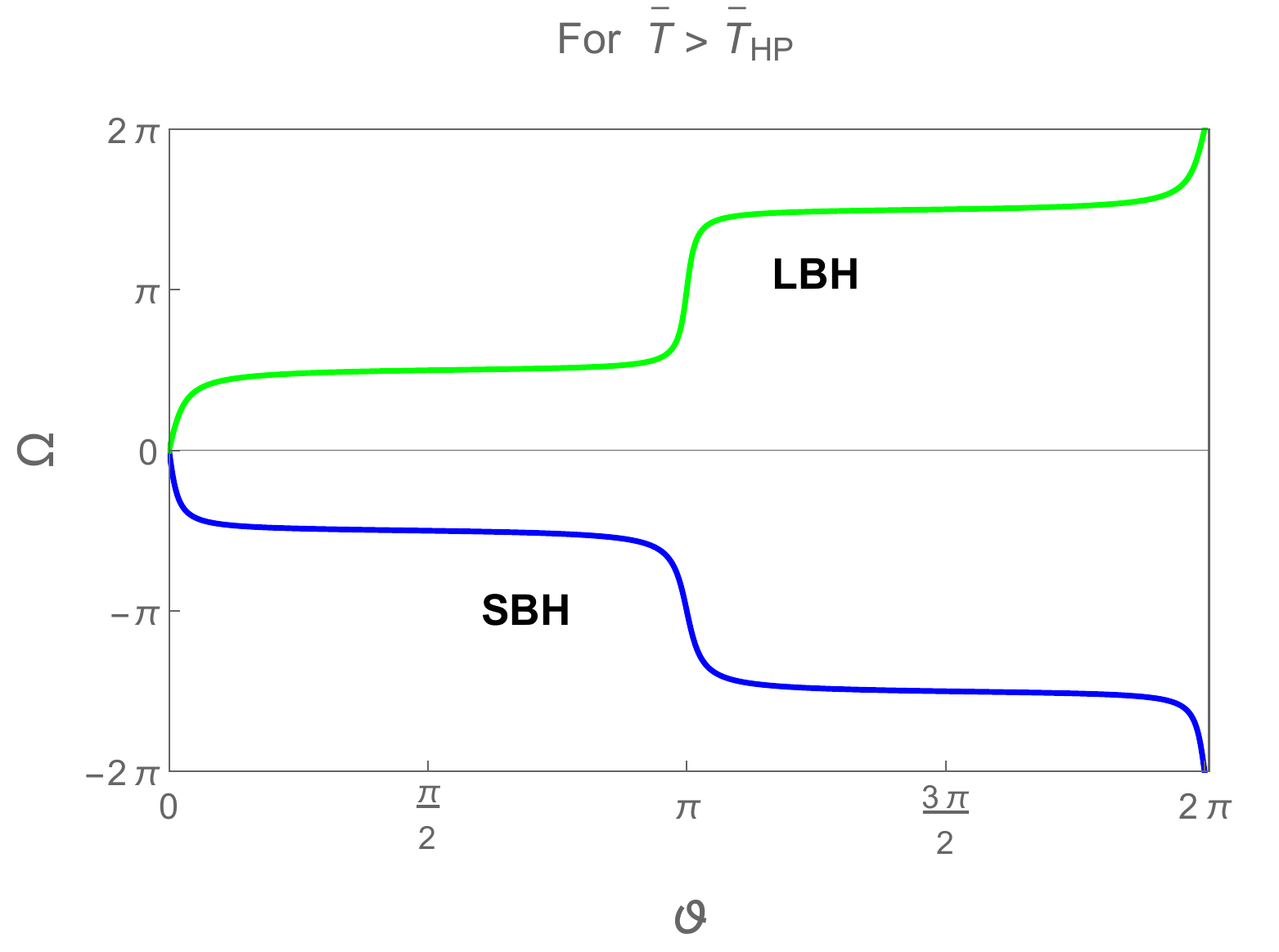}}						
		\caption{\footnotesize For Schwarzschild-AdS black hole solutions at various temperatures $\bar{T}$: Left panel shows the vanishing of vector field $\phi$ at extremal points of $\bar{f}$ (black dots) i.e., at small black holes (SBH) and large black holes (LBH) and also at the black hole possessing lowest temperature $\bar{T}_{min}$ (see plot (a)). Right panel shows the behavior of deflection angle $\Omega (\vartheta)$ for corresponding black holes in left panel. Plots are displayed for $n=2$.} 
\label{fig:8plots for sch-sol}	}
\end{figure}
\noindent
This gives,
\begin{eqnarray}
Q_c &=&  \frac{\mu (n-1)  l^{n-1} }{2\pi^2} \big(1-c^2\mu^2 \big)^{\frac{n-1}{2}} \nonumber \\
   T_c &=& \frac{n \sqrt{1-c^2\mu ^2} }{2 \pi  l} \,.
\end{eqnarray}
$T_c$ above of course matches with the Hawking-Page transition temperature written earlier in eqn.~(\ref{eq:rhp and thp_RN}).
Alternatively, one can also find $T_c$ using the first condition of eqn.~\ref{eq:cd_condition}, i.e., $W=0$ which first gives, 
\begin{equation}\label{eq: t0_boundary}
T_0= \frac{ \frac{2\pi^2 n (1-c^2\mu^2) Q}{(n-1)\mu}  + \frac{n}{l^2}\Big(\frac{2\pi^2 Q}{(n-1) \mu} \Big)^{\frac{n+1}{n-1}} }{\Big[\pi (2\pi)^{3n-2} \big(\frac{Q}{(n-1)\mu} \big)^n \Big]^{\frac{1}{n-1}}}.
\end{equation}
The confinement-deconfinement point can be located at the  minima of the curve $T_0$, as shown in the Fig.~\ref{fig:T0_bdg}.
Following methods discussed earlier, we define the vector field $\phi(\phi^{Q}, \phi^{\theta})$ as: 
\begin{eqnarray}
\phi^{Q} &=& \frac{n\Big[\pi (2\pi)^{3n-2} \big(\frac{Q}{(n-1)\mu} \big)^n \Big]^{\frac{1}{1-n}} \Bigg( (2\pi^2)^{\frac{n+1}{n-1}}\Big(\frac{Q}{(n-1)\mu} \Big)^{\frac{2}{n-1}} + 2l^2\pi^2(c^2\mu^2-1) \Bigg) \text{csc}\theta}{\mu l^2(n-1)^2}, \\
\phi^{\theta} &=& -T_{\rm 0}^{\phantom{0}} \text{cot} \theta \, \text{csc}\theta ,
\end{eqnarray}
where, $\Phi=\frac{1}{\text{sin}\theta} T_0 (Q)$.
This vector field $\phi$ vanishes exactly at the confinement-deconfinement transition point, as can be seen from the Fig.~\ref{Fig:hp_RN_bdg_vecplot}. The topological charge corresponding to this transition point turns out to be  $Q_t\big|_{\rm c}^{\phantom{c}} = \frac{1}{2\pi} \Omega (2\pi) = +1$ (given by the contour $C_1$ in Fig.~\ref{Fig:hp_RN_bdg_omegaplot}). This charge exactly matches the one obtained from the bulk calculation at the Hawking-Page transition point. 

\section{ Topological charge of equilibrium phases} \label{tcbh}

Since, the free energy and effective potential discussed in sections-(\ref{BW}) and (\ref{boundary}) respectively capture all the phases of the system, it should be possible to assign a topological charge to the phases themselves. A related idea has recently been advanced in~\cite{Wei:2022dzw},  where the black hole solutions have been identified as the topological defects (where, the vector field $\phi$ vanishes) in the thermodynamic parameter space. In this scenario, different black hole solutions have been classified into different topological classes based on the topological charge (winding) number they carry. Here we assign the topological charge (winding number) to the various phases of Schwarzschild-AdS black holes and charged-AdS black holes in grand canonical ensemble in a general situation. The extended thermodynamic set up used in~\cite{Wei:2022dzw} is though not required for the following set up.\\

\noindent
We first consider Schwarzschild-AdS black holes. We can define the vector field $\phi$, employing our Bragg-Williams free energy in eqn.~\ref{eq: BWf_sch }, as follows~\cite{Wei:2022dzw}:  
\begin{eqnarray}
\phi(\phi^{\bar{r}}, \phi^\theta) &=& \phi\Big(\frac{\partial \bar{f}}{\partial \bar{r}}, -\text{cot}\theta \text{csc}\theta \Big), \\
\text{where}, \, \phi^{\bar{r}} &=& \frac{\partial \bar{f}}{\partial \bar{r}} = \frac{n\bar{r}^{n-2} \big( n-1 +\bar{r}^2(n+1) -4\pi\bar{r}\bar{T} \big)}{16\pi}. 
\end{eqnarray}
\noindent
This vector field $\phi$ vanishes exactly at the local extremal points of $\bar{f}(\bar{r}, \bar{T})$, as shown in the Fig.~\ref{fig:8plots for sch-sol}.
Further, as we know, for $\bar{T}> \bar{T}_{min}$, the local maxima of $\bar{f}$ represent the small black holes (SBH), while its local minima  represent the large black holes (LBH). However, for $\bar{T}= \bar{T}_{min}$, only one extrema point exists for $\bar{f}$, which is neither local maxima nor minima, that represents a black hole possessing the lowest temperature $\bar{T}_{min}$. 
\vskip 0.3cm \noindent
The computation of topological charge (winding number) for these extremal points of $\bar{f}$, reveal that all the small black holes possess the topological charge $-1$, and for large black holes it is $+1$, while it vanishes for the black hole with lowest temperature (see Fig.~\ref{fig:8plots for sch-sol}). It is not difficult to check that the topological charge (winding number) associated with the charged-AdS black holes in grand canonical ensemble would be $-1/+1/0$, for SBH/LBH/black hole with lowest temperature, respectively. An analogous calculation can be set up on the boundary using the effective potential construction and is expected to give identical results.

\section{Conclusions}\label{conclusions}
In this paper, we proposed a set up to find the topological charge associated with the Hawking-Page transition point, by employing the off-shell Bragg-Williams free energy landscape used to analyze first order phase transitions.
We considered the HP transitions exhibited by Schwarzschild-AdS black hole system, and charged-AdS black hole system (in grand canonical ensemble). In both the systems, we found that the HP transition point turns out to have the topological charge $+1$. 
This is a novel topological charge, according to the classification of topological charges in~\citealp{Wei:2021vdx}. We also showed that the same value of topological charge emerges from considerations of an effective potential in the dual gauge theory, which is computed at the confinement-deconfinement transition point of the boundary gauge theory. Further it shows, first-order HP transition point and second-order  critical point in the black holes exhibiting the standard van der Waals fluid behavior, belong to different topological class~\cite{Wei:2021vdx}. 
This study opens up an important question, whether the Hawking-Page transition  simultaneously triggers the topological transition between the black hole and its background space. This requires  further study for various black hole systems and also one needs to identify the topological charge corresponding to the background space-time.    
Further, it would be interesting to find the topological charges associated with the reverse HP transitions~\cite{Mbarek:2018bau} and reentrant HP transitions~\cite{Cui:2021qpu}.
\begin{table}[h!]
\centering{		 
	\begin{tabular}{|c|c|c|c|c|c|c|}
		\hline \hline
		 & \multicolumn{3}{|c|}{} & \multicolumn{3}{|c|}{Charged-AdS BH in} \\
		  & \multicolumn{3}{|c|}{Schwarzschild-AdS BH} & \multicolumn{3}{|c|}{ grand canonical ensemble} \\   \cline{2-7} 
		&  &  &  &  &  &  \\ 
			& SBH & LBH & BH at $\bar{T}_{min}$ & SBH & LBH  & BH at $\bar{T}_{min}$ \\ \hline
		\multirow{3}{2.5cm} {Topological} & &  &  &  & &  \\ 
		& -1	& +1 & 0 & -1 & +1 & 0  \\ 
	charge (winding number)	&	& &  &  & & \\ \hline
		\multirow{3}{2.5cm} {Total} &   \multicolumn{3}{|c|}{} & \multicolumn{3}{|c|}{}   \\ 
		&	\multicolumn{3}{|c|}{0} 	& \multicolumn{3}{|c|}{0}   \\ 
 topological	charge 	& \multicolumn{3}{|c|}{} 		& \multicolumn{3}{|c|}{}  \\
		\hline\hline
	\end{tabular}
		\caption{Summary of the results.}
		\label{table:summary}}
	\end{table}
\vskip0.3cm
\noindent
We then continued to study the different equilibrium phases of the system by computing the topological charge (winding number) to the Schwarzschild-AdS black hole solutions and charged-AdS black hole solutions in grand canonical ensemble. The results are summarized in the Table~\ref{table:summary}. Our results are in accord with the conjecture of~\cite{Wei:2022dzw},  that +1/-1 topological charge (winding number) indicates the locally  stable/unstable black hole solutions.
The total topological charge (winding number) is found to be zero for both Schwarzschild-AdS black holes and charged-AdS black holes in grand canonical ensemble, which shows that these two black hole systems belong to the same topological class. Further, charged-AdS black holes in canonical~\cite{Wei:2022dzw} and in grand canonical ensemble studied here, belong to different topological classes. It would be nice to continue this topological classification of more general black holes too. Though, we note that one does not require the extended thermodynamic set up to study the topological charges of critical points.\\

\noindent
Encouraged by the fact that one is able to assign a topological charge to the deconfinement transition, we end this note with the
following question. Can we associate topological charge at the phase transition points exhibited by well known models often used for condensed matter
systems. To that effect, let us consider, for example, the Ising model in $n$ dimensions. It is expected to have a critical point at a finite
temperature, beyond which the magnetization vanishes. The BW free energy
density of the model is given in ~\cite{cha95} and has the form
\begin{equation}
f(T,m) = -\frac{1}{2} J z m^2 + \frac{1}{2} T[(1+m)~{\rm log}(1+m) + (1 - m) ~{\rm log}(1-m)] - T ~{\rm log}2.
\end{equation}
Here, $J (> 0)$ is the nearest neighbour spin-spin coupling constant, $z$ is the coordination number, $T$ and $m$ are the temperature and the magnetization 
density respectively. The critical point occurs at $T_c = Jz, m = 0$ which can be located via the equations $\partial f/\partial m = 0$
and $\partial^2 f/\partial m^2 = 0$.
To characterize this point, we define a vector field with components $\phi^m = \partial_m (T/{\rm sin}\theta)$ and
$\phi^\theta = \partial_\theta (T/{\rm sin}\theta)$. Computing the deflection angle $\Omega(2\pi)$ in an analogous manner as before, we get the topological charge to be $-1$,
if the contour encloses this critical point. Otherwise the result vanishes. The association of this charge may not provide any new information
about the model itself but leads to a novel way of characterization though.

\section*{Acknowledgements}
One of us (C.B.) thanks the DST (SERB), Government of India, for financial support through the Mathematical Research Impact Centric Support (MATRICS) grant no. MTR/2020/000135. 
\bibliographystyle{apsrev4-1}
\bibliography{topology_hp}
\end{document}